\documentclass[%
aps,prl,%
 amsmath,amssymb,
reprint,%
superscriptaddress,showpacs]{revtex4-1}

\usepackage{color, graphicx}
\usepackage{dcolumn}
\usepackage{bm}
\usepackage{amssymb}
\usepackage{latexsym}
\usepackage{amsfonts}
\usepackage{amsmath}
\usepackage{multirow}
\usepackage{ifthen}

\begin{document}

\title{Polydispersed Granular Chains: From Long-lived Chaotic Anderson-like Localization to Energy Equipartition}

\author{V.~Achilleos}
\author{G.~Theocharis}
\affiliation{ Laboratoire d'Acoustique de l'Universit\'e du Maine, UMR CNRS 6613 Av. O. Messiaen, F-72085 LE MANS Cedex 9, France}
\author{Ch.~Skokos}
\affiliation{Department of Mathematics and Applied Mathematics, University of Cape Town, Rondebosch 7701, South Africa
}

\begin{abstract}

We investigate the dynamics of highly polydispersed finite granular chains.
From the spatio-spectral properties of small vibrations, we identify which particular single-particle displacements lead to energy localization.
Then, we address a fundamental question: Do granular nonlinearities lead to chaotic dynamics and if so, does chaos destroy this energy localization?
Our numerical simulations show that for moderate nonlinearities, although the overall system behaves  chaotically, it can exhibit long lasting energy localization for particular single particle excitations. On the other hand, for sufficiently strong nonlinearities, connected with contact breaking, the granular chain
reaches energy equipartition  and an equilibrium chaotic state, independent of the initial position excitation.

\end{abstract}

\maketitle

Granular solids are densely packed assemblies of polydisperse grains commonly found in nature and industry~\cite{granular,poly,bookgran}.
Recent technical and conceptual advances on the vibrational analysis of micro~\cite{col,colglas,colcry} and macro~\cite{gran}  granular solids led to a better understanding of their dynamics and revealed novel mechanical features. 
In addition, mesoglasses made as granular assemblies of brazed aluminum beads have also been used
for fundamental studies in glass physics,
including classical (elastic) Anderson localization~\cite{Page} and its mobility gaps~\cite{Page2}.
However, to further probe the intrinsic transport and mechanical properties of granular solids, a deeper understanding of the 	\textit{anharmonic} grain contact interactions is required, including the Hertzian contact interactions, breaking of contacts and friction~\cite{bookgran}. These distinct sources of nonlinearity in combination with features like the polydispersity, make the vibrational 
energy transport in granular solids complex and an open major challenge in physics. 

A prototypical structure 
to study energy transport in the presence of granular nonlinearities
is the one-dimensional granular solid, called also \textit{granular chain}.
The study of granular chains, both homogeneous and disordered, is a vibrant area of 
research~\cite{chapter,chiaropanos,Guebelle,Luding,review}, attracting considerable attention. Nevertheless, it was only recently 
that the significance of the interplay between disorder and nonlinearity in energy transport ~\cite{MasonPanos,ourPRE} has been 
explored. Thus, further studies of the dynamics of polydispersed granular chains are essential towards a better understanding of the 
energy transport in disordered granular solids. At the same time, disordered granular chains provide a macroscopic experimentally 
accessible setup for further fundamental studies on the nonlinear dynamics of disordered media. For example, granular chains are 
ideal test bed for the study of Anderson localization in the presence of nonlinearity and the role of chaos in its long-time dynamics.

The fate of Anderson localization in the presence of nonlinearity in disordered lattice models, like the discrete Klein-Gordon 
(KG) and the discrete nonlinear Schr\"{o}dinger (DNLS) equation, 
has been intensively studied in recent years~\cite{disorder_num,disorder_other,skokflach2009,SGF13,TSL14}. 
It is now well established that nonlinear Anderson localization has probabilistic features~\cite{FlachProb}.
Depending on the initial conditions and the amount of nonlinearity, it either persists or is destroyed leading to 
energy spreading associated with chaos. For specific models in~\cite{aubryKAM} it was shown that 
the presence of initial chaos did not imply energy spreading, while in ~\cite{TSL14} it was shown that chaos is a necessary, but not sufficient, condition for thermalization. Furthermore
%
a systematic study of single site exitations in the KG model was performed in~\cite{SGF13},
where it was shown
that chaotic dynamics slows down when it is associated with energy spreading.

Aiming  to a better understanding of energy localization and transport in granular solids and to highlight the role of chaoticity, 
in this work we investigate the dynamics of a strongly polydispersed granular chain under single particle displacements.
%
Our results reveal the existence of different kinds of long time behaviors, including chaotic energy spreading, long-lived chaotic Anderson-like localization and energy equipartition.
In particular,  we report a transition from a weakly nonlinear regime featuring both diffusive transport and Anderson-like localization, which strongly depend on the initial conditions, 
to a highly nonlinear regime characterized by contact breakings and energy equipartition found to be independent of the initial conditions.

\begin{figure}
\includegraphics[width=8.5cm]{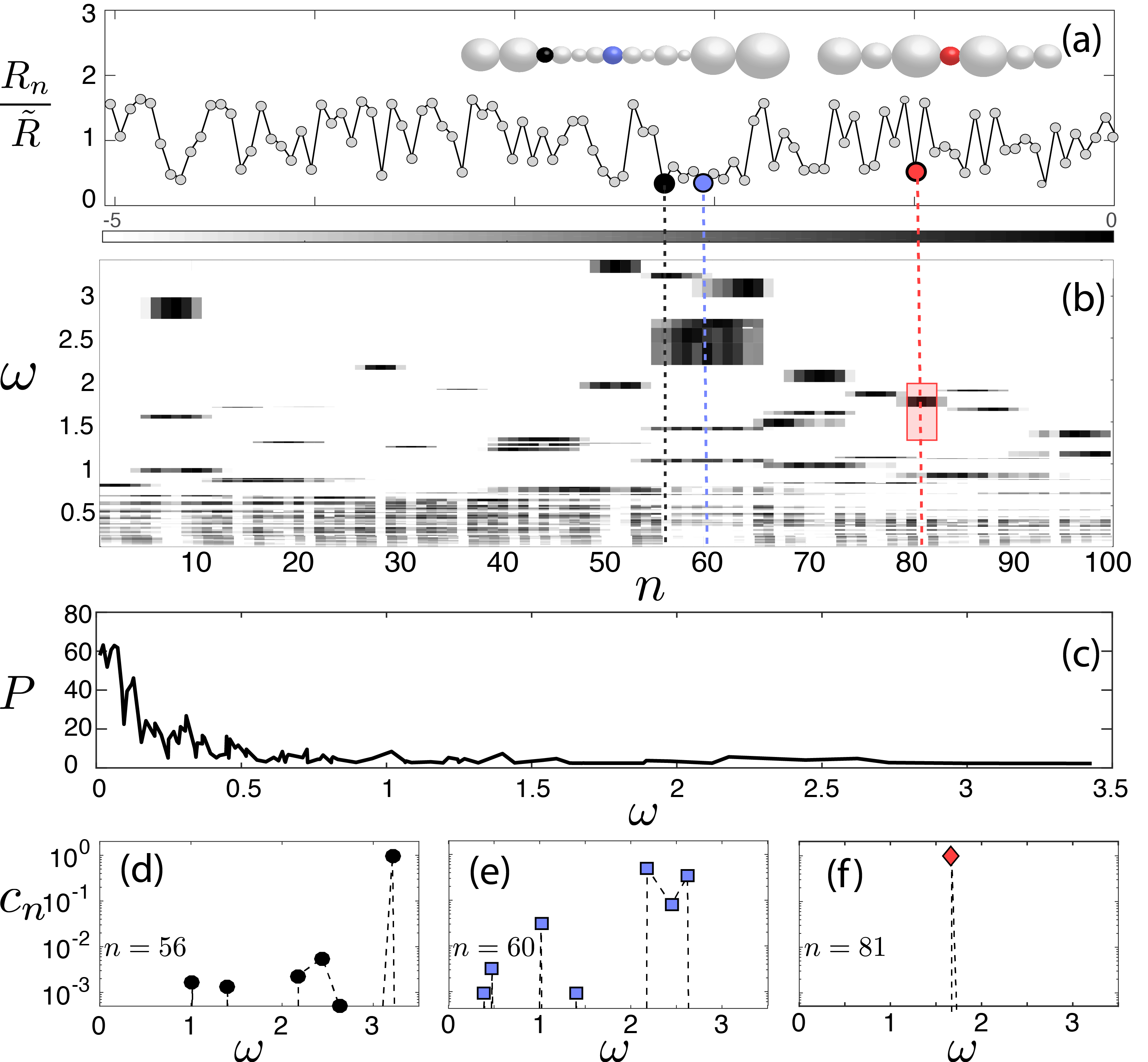}
\caption{(a) The random distribution of the normalized radii considered here. Insets show the spheres in scale, in the neighborhood of  $n=60$ and  $n=81$.
 (b) Contour plot of the coefficients ${\rm log}|c_m^n|$
 (c) The participation number $P$ of the eigenmodes as a function of their frequency.~(d)-(f) $c_m^n$ for the initial displacement of beads $n=56$, $n=60$, and $n=81$.
}
\label{sketch}
\end{figure}

The structure we study here consists of a chain of $N$ spherical particles in contact,
having masses $m_n$ ($n=1,2,\ldots, N$). The
corresponding Hamiltonian $H$  (whose value represents the system's energy)
 is given by
\begin{eqnarray}
H=\sum_{n=1}^{N} H_n\equiv \sum_{n=1}^{N}\left( \frac{p_n^2}{2m_n} +V_{n}\right),
\label{Hamlinear}
\end{eqnarray}
where $H_n$ and $p_n=m_n\dot{u}_n$ are the energy and momentum of the $n-$th spherical particle, with $u_n$ being the displacement of this particle from its equilibrium position, while
dots ($\dot{ }$) denote differentiation with respect to time.
The potential $V_n$ for each particle is defined as $V_n=[V(u_{n})+V(u_{n+1})]/2$ where:
\begin{equation}
\begin{aligned}
&V(u_n)=\frac{2}{5}A_n[\delta_n+u_{n-1}-u_n]_+^{5/2}-\frac{2}{5}A_n\delta_n^{5/2} \\
&-A_n\delta_n^{3/2}(u_{n-1}-u_n).
\end{aligned}
\label{Vhz}
\end{equation}
In Eq.~(\ref{Hamlinear}) hard wall boundary
conditions, $u_0=u_{N+1}=0$, are used.
$\delta_{n}$ is the relative static overlap due to a pre-compression force $F$ acting on the chain
and is given by $\delta_n=(F/A_n)^{2/3}$~\cite{hertzbook}, where  $A_n$ is the contact coefficient between particles $n-1, \,n$.
For spheres of the same material $A_n=(2/3)\mathcal{E}\sqrt{(R_{n-1}R_{n})/(R_{n-1}+R_n)}/(1-\nu^2)$~\cite{hertzbook},
with $\mathcal{E}$, $\nu$  and $R_n$  being the elastic modulus, the Poisson's ratio and the radius of the $n$ bead,
respectively~\cite{units}.
The terms  $[\;]_+$ in Eq.~(\ref{Vhz}) vanish when their argument
becomes negative. This happens when a \textit{gap} between two particles appears i.e.~$u_{n-1}-u_n>\delta_n$.
We normalize units using as a reference the uniform system
with particles of radius $\tilde{R}=(\alpha+1)R/2$ where $\alpha={\rm max}(R_n)/{\rm min}(R_n)$ is
the disorder strength~\cite{normalization}.

 Considering small displacements
 $u_n\ll 1$, we obtain a linearized Hamiltonian with  potential
 $V(u_n)=K_n (u_{n-1}-u_n)^2/2$,
%
%
where $K_n=(4/3)A_n\delta_n^{1/2}$~\cite{hertzbook} is the linear coupling constant.
We chose a particular disordered setup of $N=100$ spheres  whose radii  $R_n$ are taken from a uniform distribution, within the range $R_n\in [R,\alpha R]$ with  $\alpha=5$
[Fig.~\ref{sketch}(a)].

Initially we  obtain the harmonic eigenmodes and
their participation number $P=1/\sum{h^2_n}$, where $h_n=H_n/H$ 
[Fig.~\ref{sketch}(c)].   
At low frequencies, around 10 modes exist with $P>40$  exhibiting a localization length of the order of the total chain.
These are called propagons~\cite{1Dglasses}  and are responsible for the transport of energy~\cite{Kundu}
of the linearized granular chain, which is a FPU-like lattice.
The remaining modes are localized (also called locons~\cite{1Dglasses}), and among them there are at least $20$ strongly localized ones with $P\lesssim 3$, a result of high polydispersity.

In Fig.~\ref{sketch}(b) we plot the projection coefficients $c^n_m$ of all possible single site displacement excitations
(of the form $\vec{u}^{\; n}(0)=[0, 0,\ldots,1,\ldots,0]$) onto the normal modes of the harmonic chain.
The index $n$ counts the number of the bead which is displaced  [x axis of Fig.~\ref{sketch}(b)] and
$m$ counts the number of the normal modes when they are ordered in increasing frequency $\omega$ [y axis of  Fig.~\ref{sketch}(b)].
Using Fig.~\ref{sketch}(b) one is able to identify particular excitations whose coefficients  at the low frequency
extended modes, are highly suppressed. Thus, such single site excitations could provoke Anderson localization. 
Among the different choices here we focus on three particular ones, corresponding to initial
displacements of beads $n=56$, $60$ and $81$, for which in Fig.~\ref{sketch}(d),(e), and (f) we show the coefficients with $c^n_m>10^{-4}$.
{The
natural question that raises is how the granular nonlinearities  influence the aforementioned localization.
Should we also expect a chaos-induced destruction of localization and diffusive energy {spreading}
as is the case of KG and DNLS lattices?
To answer these fundamental questions, we investigate the long time dynamics of the chain
after initially displacing  particles $n=51$, $60$ and $81$.
The energy spreading is monitored by calculating the participation number $P$ and the
second energy moment $m_2=\sum_n(n-\tilde{n})^2H_n$, where $\tilde{n}=\sum nH_n$.
The system's chaoticity is quantified using the maximum Lyapunov
characteristic exponent (mLCE) $\lambda$~\cite{BGGS,S10}, which is obtained
by numerically integrating  both the Hamilton equations of motion and the corresponding variational equations~\cite{tmap} by a symplectic integration scheme~\cite{FR90_Y90}.
Note that  $\lambda$ is evaluated as $\lambda=\underset{t-\rightarrow \infty}{{\rm lim}}\Lambda(t)$, where
%
%
$\Lambda(t)$ is the so-called finite time mLCE~\cite{S10}.
For chaotic orbits $\Lambda(t)$ eventually converges to a positive value, while for regular orbits it tends to zero following the power law $\Lambda(t)\propto t^{-1}$~\cite{S10}.


%
%
\begin{figure}
\includegraphics[width=8.7cm]{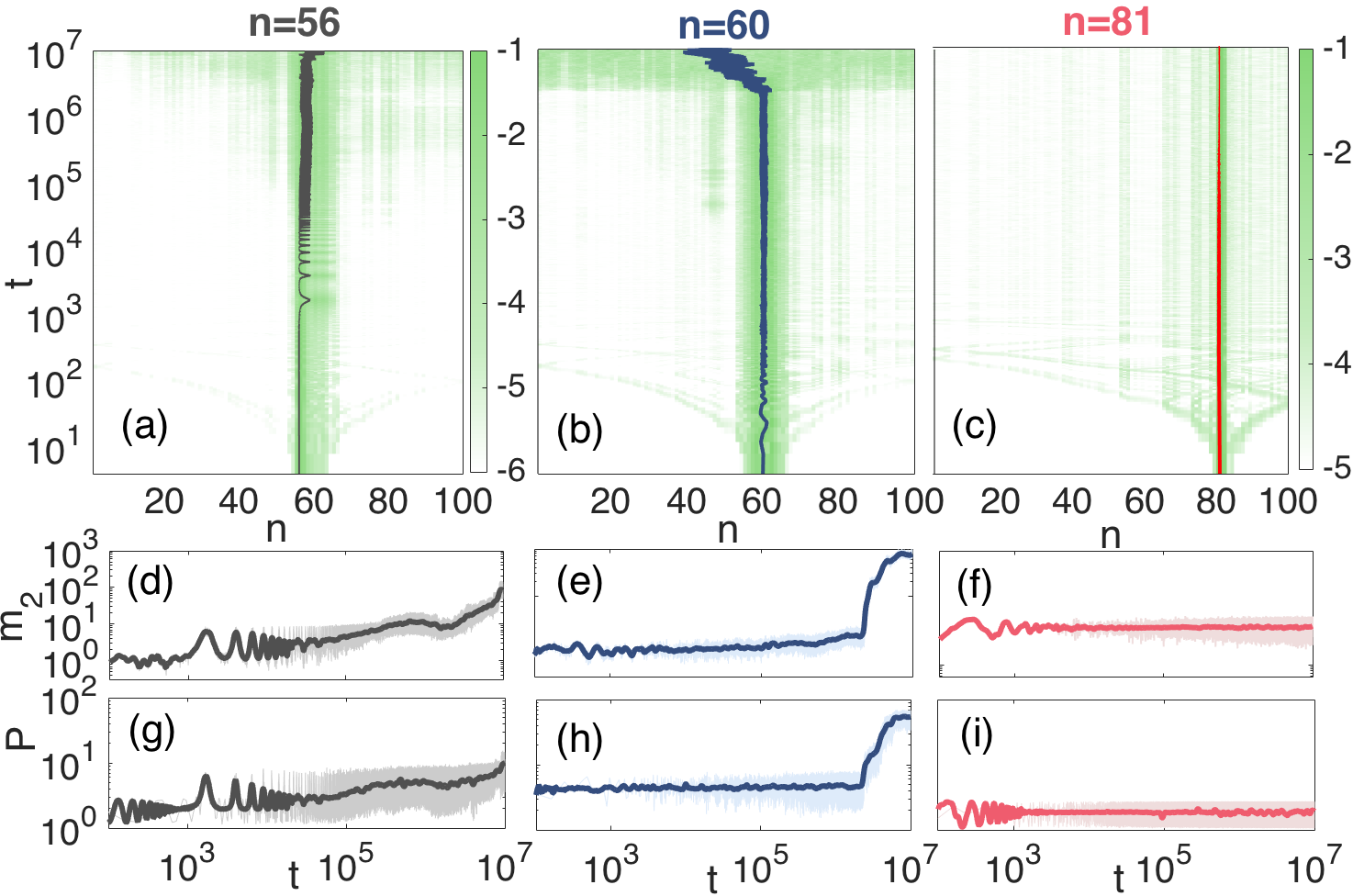}
\includegraphics[width=8.5cm]{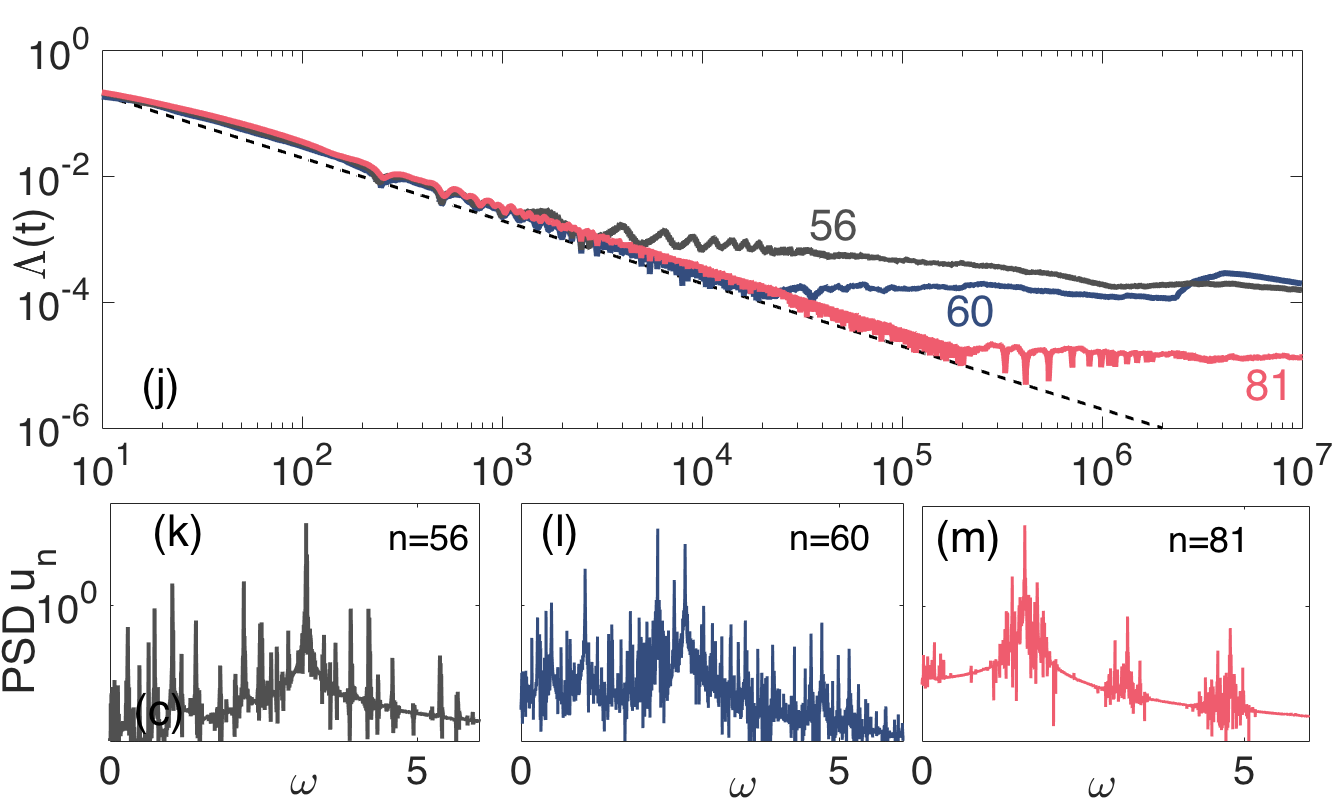}
\caption{(a)-(c) The spatiotemporal evolution of energy after exciting
bead $n=56$ (a) with energy $H=0.049$ and beads $n=60$ (b) and $n=81$ (c) with energies $H=0.192$. Color maps are
in logarithmic scale. 
The solid curves show the mean position of the energy distribution. (d)-(i) Time evolution of $m_2(t)$ [(d)-(f)] and $P$ [(g)-(i)] respectively for the cases of panels (a)-(c).  Solid curves correspond to the running averages of the plotted quantities.
(j) $\Lambda(t)$ for the three different cases of panels (a)-(c).
(k)-(m) Fourier transform taken from the momenta of beads $n=56$, $60$ and $81$ for a time interval
 of duration $\Delta t=5000$ time units after $t=4.95\times10^5$ [dashed vertical line in (j)].}
\label{linear}
\end{figure}
%
We first focus on moderate nonlinearities by imposing an initial small
amount of energy in the chain.
%
The dynamics induced by displacing bead $n=56$ with  $H=0.049$ 
is dominated by the appearance of a nonlinear coupling between three localized
modes~\cite{SM}, initially  resulting in small oscillations of the mean energy position [solid {black curve} in Fig.~\ref{linear}(a)], although eventually the wavepacket starts to spread. 
The energy transfer between the three localized modes for times $t\lesssim 10^4$
appears also as oscillations in the evolution of  $m_2$ and $P$ [Figs.~\ref{linear}(d) and (g) respectively].
At $t\approx 10^4$, $m_2$ starts increasing with a small slope, suggesting a slow spreading,
while after $t\approx 10^6$  tends to increase with an even larger slope
signaling a faster spreading  also evident in the energy profile in Fig.~\ref{linear}(a).

In Fig.~\ref{linear}(j) where we plot the corresponding $\Lambda(t)$ 
we observe that after $t\approx 10^3$ $\Lambda(t)$ deviates from  regular dynamics ($\Lambda(t) \propto t^{-1}$), a clear indication of the chaotic nature of the evolution. Nevertheless,  $\Lambda(t)$ does not remain constant 
but acquires a new slope $\Lambda(t)\propto t^{-\nu}$ where $\nu<1$.
Such a slope is associated with a non-equilibrium chaotic behavior: as the wave packet spreads exciting
more lattice sites (increasing the active degrees of freedom) the
dynamics becomes less chaotic since the constant total energy of the system is divided among more sites.
Similar behavior was also
observed in Ref.~\cite{SGF13} for a KG chain. The chaoticity of the wavepacket is also visible in the frequency spectrum of bead $n=56$
shown in  Fig.~\ref{linear}(k), as the distribution practically covers the whole frequency range.


The evolution of the energy distribution, when the bead $n=60$ is initially excited with energy $H=0.192$,
is shown in Fig.~\ref{linear}(b) and the wavepacket appears to remain localized at least up
to $t\approx10^6$. Moreover, 
since $m_2$ in Fig.~\ref{linear}(e) is increasing with a very small slope, we conclude that the wavepacket is spreading very
slowly up to $t\approx 10^6$. For the same time interval, $P$  remains practically constant [Fig.~\ref{linear}(h)].
After $t\approx 10^6$, the wavepacket spreads abruptly and in the last decade of its evolution
it appears to have spread throughout the whole chain.
%
At the same time $P$ has taken a maximum value of $P\approx 60$ [Fig.~\ref{sketch}(c)].
Similar abrupt spreadings have been previously observed in KG and DNLS models~\cite{skokflach2009}.

The evolution of the corresponding $\Lambda(t)$  in Fig.~\ref{linear}(j)
indicates a primal phase of regular behavior up to $t\approx 2\times 10^4$ when $\Lambda(t)$
starts deviating from the $\propto t^{-1}$ law. Then, during the very slow spreading phase which follows for  $2\times 10^4 \lesssim t \lesssim 2\times 10^6$, $\Lambda(t) \approx 10^{-4}$.
This behavior of $\Lambda(t)$ indicates that the weakly spreading phase is connected to a slow
thermalization process induced by the chaotic response and the corresponding  chaotic dephasing of the initially excited normal modes.
However, following the abrupt spreading appearing in the evolution of the energy distribution in the
second column of Fig.~\ref{linear}, $\Lambda(t)$ exhibits a `jump' to  higher values. This jump is related to the transition of
the motion from a `small chaotic sea'  which is confined in a small
subset of the system's phase space, i.e.~the motion of  few sites around the bead $n=60$, to a `large chaotic sea' which occupies almost all phase space, i.e.~the chaotic motion of  the whole chain (a similar transition was reported in \cite{konto78}). The chaotic nature of the wavepacket is also reflected in the almost continuous frequency spectrum of Fig.~\ref{linear}(l).
In Fig.~\ref{linear}(c), we  show the spatio-temporal evolution of the energy after exciting bead $n=81$ with $H=0.192$. Here we have a unique behavior since, contrary to the two other cases,
we do not observe a destruction of the energy localization, at least for the considered integration times.
Consequently, both $m_2$ and $P$ remain practically constant [Figs.~\ref{linear}(f) and (i)].
In fact, according to Figs.~\ref{linear}(c) and (i) only beads $n=80$-$82$ significantly contribute in the dynamics
of the system since $P\approx 2$.
What is more interesting in this case, is the evolution of $\Lambda(t)$  in Fig.~\ref{linear}(j), as
it deviates from the $\propto t^{-1}$ law at $t \approx 2 \times 10^5$}  and
appears to remain positive and practically constant until the end of the integration.
This behavior indicates that the motion of the localized wavepacket is chaotic. {In addition, the constancy of
$\Lambda(t)$ for $t\gtrsim 2 \times 10^5$} stems from the fact that  no
more degrees of freedom are activated  at least up {to $t=10^7$}.
Thus, for this particular case, chaoticity is not sufficient to induce the spreading of the wavepacket.
We note here that eventually this chaotic response could lead to energy spreading through
very slow processes such as Arnold {diffusion~\cite{arnold64,lieberman92,note1},  but to
the best of our knowledge this is the first time that such a long lasting spatially confined
chaos is reported in a disordered, non-degenerate, lattice system.
From the corresponding frequency spectrum  in
Fig.~\ref{linear}(m) we observe that only a finite range of frequencies located around
the fundamental ones and their higher harmonics are excited. According to the spatio-spectral properties of the linear system,
the evolving beads $n=80$-$82$, do not participate to linear modes within this range of frequencies [see
the red square of Fig.~\ref{sketch}(b)]. This qualitatively explains the robustness of this Anderson-like localization despite the chaotic nature of the wavepacket.

\begin{figure}
\includegraphics[width=8.5cm]{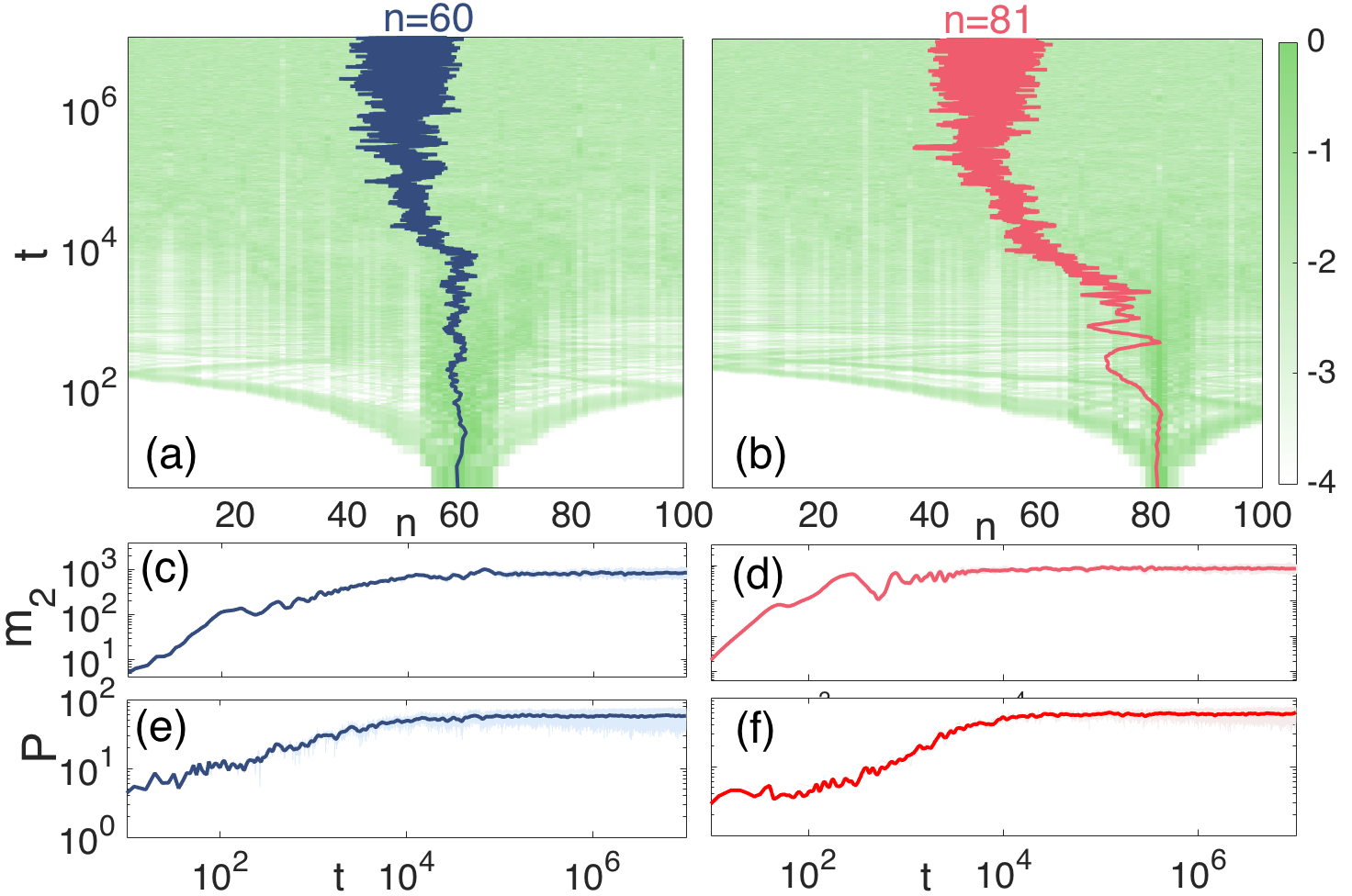}
\includegraphics[width=8.5cm]{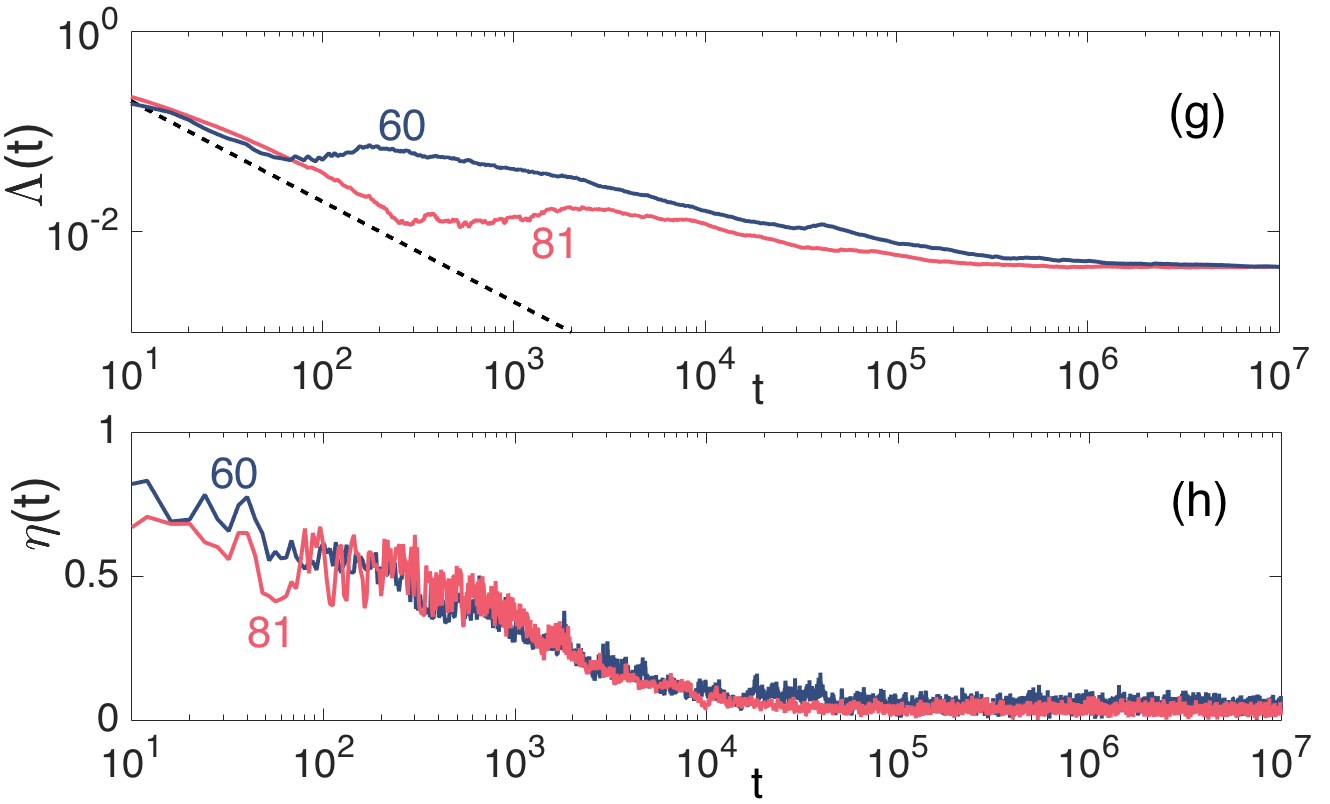}
\caption{(a) and (b): Same as in Fig.~\ref{linear} (b)-(c)
but with energy $H_1=2.76$.
(c)-(d) Same as in Fig.~\ref{linear} (e)-(h) and  (e)-(f) Same as in Fig.~\ref{linear} (f)-(i).
(g) $\Lambda(t)$  for the  cases of panels (a) and (b). (h) The time evolution of the normalized spectral entropy $\eta(t)$ (\ref{norment}) for the cases of panels (a) and (b).
}
\label{hnl}
\end{figure}
%

We furthermore,  study the dynamics of the chain after single particle displacements with energies an order of magnitude
larger than then ones of Fig.~\ref{linear} to investigate the effect of stronger nonlinearites.
Characteristic examples of the energy evolution in this regime are shown in Figs.~\ref{hnl}(a) and (b),
after the initial displacement of particles $n=60$ and $81$ respectively, with $H=2.76$.
It is directly evident from Fig.~\ref{hnl}(a) and (b), that both wavepackets remain initially localized
but finally spread throughout the whole chain, as is also indicated by the mean position of the energy distribution
which finally oscillates around the chain's center for $t\gtrsim 10^4$.
{The evolution of $m_2$ is quite similar for the two cases
and is found to saturate around the same value $m_2\approx 10^3$ [Figs.~\ref{hnl}(c) and (d)].
Similarly,  $P$  saturates to a constant value around  $P\approx 60$ for both excitations [Figs.~\ref{hnl}(e) and (f)].}
Note that in contrast to the case of weak excitations, here we observe
the opening of gaps that travel through the whole chain~\cite{SM}.

Regarding the chaoticity in this regime,  $\Lambda(t)$ shown in Fig.~\ref{hnl}(g), indicates
a clear chaotic behavior 
clearly deviate from the  $t^{-1}$ law [dashed line of Fig.~\ref{hnl}(g) after $t\approx 10^2$ ]. 
Furthermore, for both cases, after about $t\approx 10^6$, $\Lambda(t)$
saturates to almost the same  constant, positive value. 
The evolution of $\Lambda(t)$ for $10^2 \lesssim t \lesssim 10^6$ suggests
a transient chaotic behavior for the chain, where
energy spreading takes place and more degrees of freedom are activated. Within this time window
the number of gaps appearing in the system is changing, and remain constant only after $t \gtrsim 10^5$~\cite{SM}.

The fact that the mean values of $m_2$,  $P$, the value of $\Lambda$ [Fig.~\ref{hnl}]
and  the number of gaps in the chain~\cite{SM} saturate to the same values for both excitations
%
%
%
suggests that the system finally reaches an equilibrium state characterized by equipartition of energy.
To verify this assumption, we use the spectral entropy $S(t)$ defined by~\cite{FPUEqui}:
\begin{eqnarray}
S(t)=-\sum_{k=1}^{N} w_k(t)\,  \ln\left( w_k(t) \right),
\label{spectralentr}
\end{eqnarray}
where  $w_k$ are  weights given by the fraction of the total
harmonic energy $E_k$ in the $k$-th normal mode, {i.e.~$w(k)=E_k/ \sum_{k=1}^N E_k$, or} more conveniently the normalized
spectral entropy
 \begin{eqnarray}
\eta(t)=\frac{S(t)-S_{max}}{S(t)-S{(0)}},
\label{norment}
\end{eqnarray}
where $S_{max}$ is the maximum measured entropy.
$\eta=1$ 
indicates that all the energy remains in the initially excited modes,
while $\eta=0$ {denotes} equipartition. From the evolution of $\eta(t)$  in {Fig.~\ref{hnl}(h)}
we observe that after some relaxation time $\eta(t)$ saturates to a value $\eta \approx 0.01$ for both initial excitations.
This behavior strongly supports the assumption that a final equilibrium  state, 
characterized by energy equipartition with the same statistical characteristics for any initial condition with the same total energy, is reached.
Importantly, this is not the case in the weakly nonlinear regime where only for some initial excitations at a given energy an equipartition is reached, similar to what was observed in~\cite{Mula}.

%



In this work we studied the chaotic behavior of a finite, strongly disordered granular chain for single bead initial
displacements. Based on the spatio-spectral properties of the
corresponding harmonic chain we were able to identify excitations which lead, for moderate energies, to long-lived, chaotic Anderson-like localized motion. In addition, we showed that for sufficiently strong initial excitations the coexistence of anharmonic nearest neighbors nonlinearities and gaps  lead to a chaotic destruction of localization, and the system finally reaches the same equilibrium state of energy equipartition for all initial conditions with the same energy.

G.T.~acknowledges financial support from FP7-CIG (Project 618322 ComGranSol). Ch.S.~acknowledges support by the National Research Foundation
of South Africa (IFRR and CPRR Programmes) and thanks LAUM for its
hospitality during his visit when part of this work was carried out.


\begin{thebibliography}{99}

\bibitem{granular}
H.M.~Jaeger and S.R.~Nagel,  Rev. Mod. Phys. \textbf{68}, 1259 (1996).

\bibitem{poly}
J. K. Mitchell and K. Soga, Fundamentals of Soil Behavior, 3rd ed. (Wiley, New York, 2005); T. Aste and D. Weaire, The
Pursuit of Perfect Packing (Institute of Physics, Bristol, 2000).
%

\bibitem{bookgran} B. Andreotti, Y. Forterre and O. Pouliquen, Granular media: between fluid and solid (Cambridge University Press, New York, 2013).

\bibitem{col}
P.J. Yunker, K. Chen, M.D. Gratale, M.A. Lohr, T. Still and A.G. Yodh,  Rep. Prog. Phys. \textbf{77}, 056601 (2014).

\bibitem{colglas}
D. Kaya, N.L. Green, C.E. Maloney, M.F. Islam,  Science, \textbf{329}, 656 (2010); K. Chen, W.G. Ellenbroek, Z. Zhang, D.T. N. Chen, P.J. Yunker, S. Henkes, C. Brito, O. Dauchot, W. van Saarloos, A.J. Liu, and A. G. Yodh, Phys. Rev. Lett. \textbf{105}, 025501 (2010); A. Ghosh, V.K. Chikkadi, P. Schall, J. Kurchan, and D. Bonn, Phys. Rev. Lett. \textbf{104}, 248305 (2010).

\bibitem{colcry}
K. Chen,T. Still, S. Schoenholz, K.B. Aptowicz,
M. Schindler, A.C. Maggs, A.J. Liu, and A. G. Yodh, Phys. Rev. E \textbf{88}, 022315 (2013).

\bibitem{gran}
E.T. Owens and K.E. Daniels,  Soft Matter, \textbf{9}, 1214 (2013).

\bibitem{Page} H. Hu, A. Strybulevych, J. H. Page, S. E. Skipetrov,  and B. A. van Tiggelen,
Nature Phys. {\bf 4}, 945 (2008).

\bibitem{Page2} L. A. Cobus, S. E. Skipetrov, A. Aubry, B. A. van Tiggelen,  A. Derode, and J. H. Page, 
Phys. Rev. Lett., \textbf{116}, 193901 (2016).


\bibitem{chapter}
G. Theocharis, N. Boechler, C. Daraio, \textit{Acoustic Metamaterials and Phononic Crystals},
Springer Series in Solid-State Sciences, \textbf{173}, 217 (2013).

\bibitem{chiaropanos}
L. Ponson, N. Boechler, Y. M. Lai, M. A. Porter, P. G. Kevrekidis, and C. Daraio,
Phys. Rev. E \textbf{82}, 021301 (2010).

\bibitem{Guebelle}
Mohith Manjunath, Amnaya P. Awasthi, and Philippe H. Geubelle, Phys. Rev. E, {\bf 85}, 031308 (2012).

\bibitem{Luding}
Brian P. Lawney, Stefan Luding, Acta Mchanica \textbf{225}, 2385 (2014).

\bibitem{review}
C. Chong, M.A. Porter, P.G. Kevrekidis, C. Daraio, Nonlinear Coherent Structures in Granular Crystals, arXiv:1612.03977v1


\bibitem{MasonPanos} A. J. Mart\'{i}nez, P. G. Kevrekidis, M. A. Porter, Phys. Rev. E \textbf{93}, 022902 (2016).

\bibitem{ourPRE} V. Achilleos, G. Theocharis, Ch. Skokos,
Phys. Rev. E \textbf{93}, 022903 (2016).





\bibitem{disorder_num}
G.~Kopidakis, S.~Komineas, S.~Flach and S.~Aubry, Phys. Rev. Lett., \textbf{100}, 084103 (2008); A.S.~Pikovsky  and D.L.~Shepelyansky, Phys. Rev. Lett., \textbf{100},  094101 (2008); S.~Flach, D.O.~Krimer and Ch.~Skokos, Phys. Rev. Lett., \textbf{102}, 024101 (2009); I. Garc\'{i}a-Mata and D.L.~Shepelyansky, Phys. Rev. E \textbf{79},
026205 (2009); T.V.~Laptyeva, J.D.~Bodyfelt,  D.O.~Krimer, Ch.~Skokos and S.~Flach, EPL, \textbf{91}, 30001 (2010); Ch.~Skokos and S.~Flach, Phys. Rev. E, \textbf{82}, 016208 (2010); S.~Flach, Chem. Phys., \textbf{375}, 548 (2010); J.D.~Bodyfelt, T.V.~Laptyeva, Ch.~Skokos, D.O.~Krimer and S.~Flach, Phys. Rev. E, \textbf{84}, 016205 (2011); Ch.~Antonopoulos, T.~Bountis, Ch.~Skokos and L.~Drossos,  Chaos, \textbf{24}, 024405 (2014).

\bibitem{disorder_other}
H.~Veksler, Y.~Krivolapov  and S.~Fishman, Phys. Rev. E, \textbf{80}, 037201 (2009);  D.~Basko, Ann. Phys. (N.Y.), \textbf{326}, 1577 (2011); J.D.~Bodyfelt, T.V.~Laptyeva., G.~Gligoric, D.O.~Krimer, Ch.~Skokos and S.~Flach, Int. J. Bifurcat. Chaos, \textbf{21}, 2107 (2011); M.~Mulansky and A. Pikovsky, Phys. Rev. E, \textbf{86}, 056214 (2012); T.V.~Laptyeva, J.D.~Bodyfelt, and S.~Flach, Europhys.
Lett. \textbf{98}, 60002 (2012); M.~Mulansky and A. Pikovsky, New J. Phys. \textbf{15}, 053015 (2013); M.V.~Ivanchenko, T.V.~Laptyeva and S.~Flach, Phys. Rev. B, \textbf{89}, 060301(R) (2014); T.V.~Laptyeva, M.V.~Ivanchenko and S.~Flach, J. Phys. A: Math. Theor. \textbf{47} 493001 (2014).

\bibitem{skokflach2009}
 Ch.~Skokos, D.O.~Krimer, S.~Komineas and S.~Flach, Phys. Rev. E, \textbf{79}, 056211 (2009);.

\bibitem{TSL14}
O.~Tieleman, Ch.~Skokos and A.~Lazarides A., Europhys.
Lett., \textbf{105}, 20001 (2014).

\bibitem{SGF13}
Ch.~Skokos, I.~Gkolias and S.~Flach, Phys. Rev. Let., \textbf{111}, 064101 (2013).


\bibitem{FlachProb}
M.V.~Ivanchenko,T.V.~Laptyeva, and S.~Flach, Phys. Rev. Lett. \textbf{107} 240602 (2011). 

\bibitem{aubryKAM}
M.~Johansson, G.~Kopidakis and S.~Aubry, Europhys.
Lett., \textbf{91}, 50001 (2010); S. Aubry, Int. J. Bifurcation Chaos \textbf{21}, 2125 (2011).


\bibitem{hertzbook}
 K.L. Johnson,  \textit{Contact Mechanics} (Cambridge Univ. Press, 1985).
V.F.  Nesterenko, \textit{ Dynamics of Heterogeneous Materials} (Springer, 2001)

\bibitem{units}
In our simulations we choose units corresponding to a mean radius of $R=0.01$~m,
 and a static force $F=1$~N. The elastic modulus  is chosen as $\mathcal{E}=193$ GPa and the Poisson ratio
 is $\nu= 0.3$ relevant to stainless steel (316 type).

\bibitem{normalization}
Then time, distance, mass and stiffness are scaled as follows
$t\rightarrow \tilde{\omega}t$, $\delta_n\rightarrow\delta_n/\tilde{\delta}$ $(u_n\rightarrow u_n/\tilde{u})$,
 $m_n\rightarrow m_n/\tilde{m}$, $ A_n\rightarrow A_n/6\tilde{A}$
where all the quantities with tilde, are calculated at $\tilde{R}$.
Normalization is such that  in the case of no disorder ($\alpha=1$) the normalized linear cutoff frequency
is $\omega=1$.

\bibitem{1Dglasses}
J. Fabian,  Phys. Rev. B, \textbf{55}, R3328 (1997); D.M.~Leitner, Phys. Rev. B, \textbf{64}, 094201 (2001).

\bibitem{Kundu} P. K. Datta and K. Kundu, Phys. Rev. B {\bf 51}, 6287 (1995).

\bibitem{BGGS}
G.~Benettin, L.~Galgani, A.~Giorgilli and J.-M.~Strelcyn, Meccanica, \textbf{15}, 9 (1980); ibid. \textbf{15}, 21 (1980).

\bibitem{S10}
Ch.~Skokos, Lect. Notes Phys., \textbf{790}, 63 (2010).

\bibitem{tmap}
Ch.~Skokos and E.~Gerlach,  Phys. Rev. E, \textbf{82}, 036704 (2010); E.~Gerlach, and Ch.~Skokos, Discr.Cont. Dyn. Sys.-Supp. 2011, 475; E.~Gerlach, S.~Eggl and Ch.~Skokos, Int. J. Bifurcat. Chaos, \textbf{22}, 1250216 (2012).

\bibitem{FR90_Y90}
\'{E}.~Forest and R.D.~Ruth, Physica D \textbf{43}, 105 (1990); H.~Yoshida, Phys. Lett. A,  \textbf{150}, 262 (1990).

\bibitem{SM} See Supplemental Material at [URL will be inserted by publisher]


\bibitem{konto78}  G. Contopoulos, L. Galgani, and A. Giorgilli, Phys. Rev. A \textbf{18}, 1183 (1978).

\bibitem{arnold64}  Arnold, V. I.  Sov. Math. Dokr. \textbf{5}, 581 (1964).
\bibitem{lieberman92} Liechtemberg and Lieberman, \textit{Regular and chaotic dynamics}, Springer-Verlag (1992).

\bibitem{note1}
Arnold diffusion in  systems with more than 2
degrees of freedom can allow  chaotic motion  to propagate to phase  space regions far away from the motion's origin through  a web
of (possible infinitesimally thin) interconnected chaotic layers.

\bibitem{FPUEqui}
L. Casetti, M. C-Sola, M. Pettini, and E. G. D. Cohen, Phys. Rev. E., \textbf{55}, 6566 (1997).

\bibitem{Mula}
M. Mulansky, K. Anhert, A. Pikovsky, and D. L. Shepelyansky, Phys. Rev. E., \textbf{80}, 056212 (2009).

%
%
%
%
%
%
%
%
%
%
%
%
%
%
%
%
%
%
%
%
%
%
%
%
%
%
%
%
%
%
%
%
%
%
%
%
%
%
%
%
%
%
%

\end{thebibliography}
\end{document}